\documentclass[fleqn,10pt, twocolumn]{wlscirep}
\usepackage[utf8]{inputenc}
\usepackage[T1]{fontenc}
\usepackage{lineno}
\title{Pulse Sequence Design for High Field NMR with NV Centers in Dipolarly Coupled Samples}

\author[1,2,3,*]{Carlos Munuera-Javaloy}
\author[2,4,5]{Ander Tobalina}
\author[1,2]{Jorge Casanova}
\affil[1]{Department of Physical Chemistry, University of the Basque Country UPV/EHU, Apartado 644, 48080 Bilbao, Spain}
\affil[2]{EHU Quantum Center, University of the Basque Country UPV/EHU, Leioa, Spain}
\affil[3]{Institut f\"ur Theoretische Physik und IQST, Albert-Einstein-Allee 11, Universit\"at Ulm, D-89081 Ulm, Germany}
\affil[4]{Department of Applied Mathematics, University of the Basque Country UPV/EHU, 01006 Vitoria-Gasteiz, Spain}
\affil[5]{Arquimea Research Center, 38320 La Laguna, Spain}

\affil[*]{carlos.munuera-javaloy@uni-ulm.de}

\begin{abstract}
Diamond-based quantum sensors have enabled high-resolution NMR spectroscopy at the microscale in scenarios where fast molecular motion averages out dipolar interactions among target nuclei. However, in samples with low-diffusion, ubiquitous dipolar couplings challenge the extraction of relevant spectroscopic information. In this work we present a protocol that enables the scanning of nuclear spins in dipolarly-coupled samples at high magnetic fields with a sensor based on nitrogen vacancy (NV) ensembles. Our protocol is based on the synchronized delivery of radio frequency (RF) and microwave (MW) radiation to eliminate couplings among nuclei in the scanned sample and to efficiently extract target energy-shifts from the sample's magnetization dynamics. In addition, the method is designed to operate at high magnetic fields leading to a larger sample thermal polarization, thus to an increased NMR signal. The precision of our method is ultimately limited by the coherence time of the sample, allowing for accurate identification of relevant energy shifts in solid-state systems.
\end{abstract}
\begin{document}

\flushbottom
\maketitle
%
%
\thispagestyle{empty}

\section*{Introduction}

Over the past decade, quantum sensing--a notably prolific branch of quantum technologies~\cite{Dowling03, Degen17}, has produced magnetometers able to detect ever weaker fields. Such development has had a deep impact in the realm of nuclear magnetic resonance (NMR) \cite{Levitt08}, a field that, despite its unquestionable success, has  limitations due to the inherent weakness of target signals  necessitates the scanning of millimeter-sized samples. Nitrogen vacancy (NV) centers in diamond \cite{Doherty13}, however, reported the detection of signals from smaller samples leading to NMR experiments with unprecedented spatial resolution. NV centers stand out for their capacity to operate at room temperature leading to smaller, and easier to operate magnetometers compared to platforms that require stringent conditions such as, e.g., superconducting quantum interference devices (SQUIDs).

NMR spectroscopy reveals frequency shifts relative to a base Larmor frequency which encode structural information about sample molecules as well as of their surrounding environment. In this scenario, one of the most impressive results produced by NV based NMR sensors --enabled by heterodyne protocols that overcome the resolution boundary posed by the coherence time of the NV centers~\cite{Boss17,Schmitt17}-- is the record of spectral features from picoliter volume samples with high-resolution~\cite{Glenn18}. Nevertheless, in thermally polarized samples, these protocols require a high nuclear spin concentration (typically, highly protonated samples) and a significant number of repetitions to obtain meaningful results. A possibility for detecting samples at lower concentrations and achieving more competitive protocols is to hyperpolarize the samples in a previous step~\cite{Arunkumar21, Bucher20}. In addition, performing the experiment at high fields directly provides higher polarization rates while it facilitates the extraction of relevant information from the recorded spectra, as chemical shifts increase and J-couplings become clearer~\cite{Ugurbil03}. Although NV-NMR spectroscopy at high fields has not yet been experimentally demonstrated, recent proposals have introduced protocols that enable the acquisition of high-resolution spectra in strong external magnetic fields \cite{Munuera23, Meinel23,Alsina23}. In particular, our proposal AERIS \cite{Munuera23} operates encoding the target nuclear energy shifts in the amplitude variation of the sample's longitudinal magnetization which oscillates at a tunable rate --i.e., at a slow rate of, typically, tens of KHz even at large fields-- during consecutive detections.

\begin{figure*}[t]
\centering
\includegraphics[width= .85\linewidth]{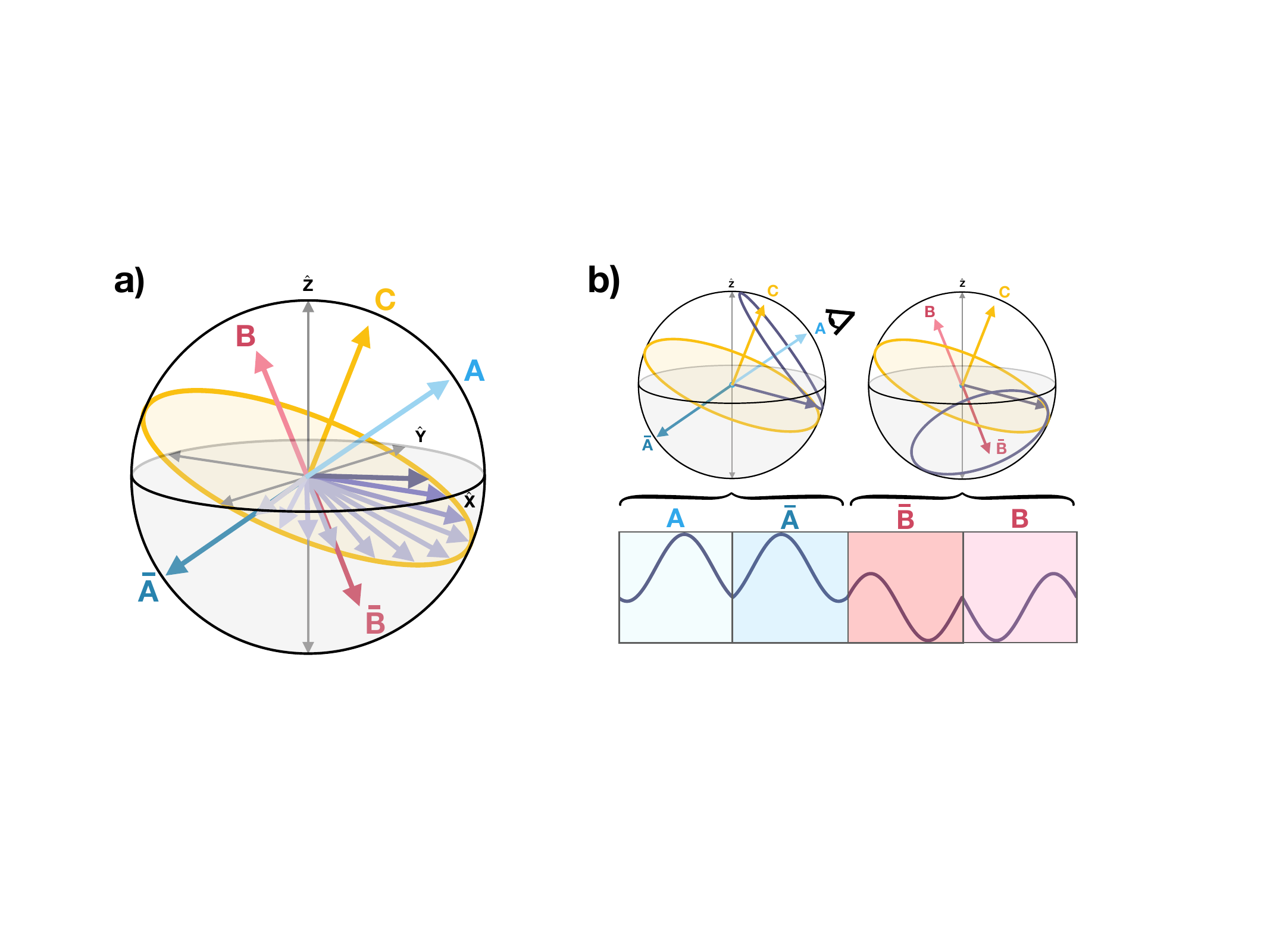}
\caption{\label{rotations} (a) Relative positions of axis $A$ (clear-blue), $\bar{A}$ (dark-blue), $B$ (clear-magenta), $\bar{B}$ (dark-magenta). 
The motion of the magnetization inbetween LG4 blocks ($A\bar A\bar B B$) is shown in purple. This evolution is described as a rotation in the plane (in yellow) perpendicular to the $C$ axis in accordance with the effective Hamiltonian in Eq.~\eqref{eq: Heff} for a single $\delta_i^*$. (b) (Top) Magnetization rotations during each of the four RF drivings. (Bottom) Projection of the magnetization onto $\hat{z}$ as it rotates during a full LG4 block. This projection determines the magnetic signal for the NV ensemble sensor.}
\end{figure*}

An important limitation of state-of-the-art NV-NMR spectroscopy is caused by the strong dipolar coupling among target nuclei, which results in intricate spectra that challenge data interpretation. This becomes especially critical in solid-state material research, where homonuclear dipole-dipole interactions hinder subtler couplings (heteronuclear interactions are less challenging as they can be eliminated by driving only one of the species). The NMR community has dedicated significant efforts to mitigate the impact of dipolar interactions \cite{Madhu16}, and today solid-state NMR spectroscopy is extensively used in distinct research areas: In pharmaceutics it characterizes active pharmaceutical ingredients (APIs) and their interaction with excipients (inactive substances added to a drug that serve various purposes such as binding or preserving the API) \cite{Pisklak14,Pisklak16,Nie16}, among many other applications (see \cite{Marchetti21} for an extensive review on the topic). In epidemiology, it provides key insights of the structure of molecules related with diseases as present in our societies as Alzheimer \cite{Tycko11}. In energy storage research is used to characterize the local structure of solid materials used in batteries and fuel cells \cite{Pecher17}.

These areas, along with many others, would largely benefit from sensors able to produce narrower spectral lines from smaller solid-state samples, especially over material surfaces where conventional NMR techniques are highly constrained. While NV-NMR spectroscopy emerges as the prominent technique to access samples in the microscale regime, it is currently limited to liquid state samples where dipole-dipole interactions get naturally averaged out due to fast molecular motion, leaving a plethora of solid-state applications out of its range of action. Extending high-field NV-NMR protocols, such as AERIS, to solid-state samples by effectively decoupling dipolar interactions, would unlock these applications \cite{Allert22, Rizzato23}.

In this work, we devise a protocol that overcomes these limitations and enables high resolution NV-NMR spectroscopy of single crystals with strong homonuclear dipolar coupling at elevated external magnetic fields, allowing to take advantage from the higher polarization rates and stronger chemical shifts.  Our protocol features the delivery of two radiation channels --radio-frequency (RF) and microwave (MW)-- synchronized with measurements on an NV ensemble magnetometer. The RF channel drives the sample with a twofold purpose. On the one hand it effectively decouples the target nuclear spins, diminishing the effect of strong homonuclear couplings in the recorded spectra, and enabling the obtention of nuclear energy shifts. Remarkably, the decoupling benefits from increasing RF intensities could be specially effective in the small volume regime, where the driving field tends to be more homogeneous. In this regime, current RF antenna designs have demonstrated nuclear spin rotation rates in the tens to hundreds of kilohertz range \cite{Herb20, Yudilevich23}. On the other hand, the RF bridges the interaction among NV sensors and fast rotating nuclear spin by generating a slow-frequency NMR signal trackable by the NV sensor. Simultaneously, the MW channel delivers a tailored pulse sequence to the NV ensemble enabling the detection of the magnetic field emitted by the driven sample. This sequence is interspersed with measurements of the sensor's state to construct the spectra in a heterodyne frame leading to a spectrum only limited by the nuclear sample coherence. Finally, we provide analytical expressions that map the detected resonances with target energy shifts.

\section*{Methods}
\subsection*{RF modulation of nuclear spins}
Lee and Goldburg (LG) showed in a seminal paper~\cite{Lee65} that an off-resonant continuous RF field cancels, up to first order, the contribution to the nuclear spin dynamics of homonuclear dipole-dipole interactions if the {\it LG condition} $\Delta=\pm \Omega/\sqrt{2}$ holds. Here, $\Delta = \omega_L-\omega_d$ is the detuning between the carrier frequency of the RF driving field ($\omega_d$) and the Larmor precession of the spins ($\omega_L$), and $\Omega$ is the Rabi frequency of the RF driving. Subjecting a spin ensemble to an off-resonant RF field leads to collective nuclear spin rotations along an axis tilted with respect to $\hat{z}$ (the direction of the static magnetic field). More specifically, {\it the tilted axis} --in the following $P$-- has a component $\frac{\Delta}{\sqrt{ \Omega^2 + \Delta^2}}$ along $\hat{z}$, while its projection on the orthogonal $xy$ plane is $\frac{\Omega}{\sqrt{ \Omega^2 + \Delta^2}}$.

Further developments have built upon the original LG sequence demonstrating the ability to remove higher order contributions of the dipole-dipole interaction, thus leading to even narrower spectral lines. Prominent examples are the frequency-switched (FSLG)~\cite{Waugh72} and phase-modulated (PMLG)~\cite{Vinogradov99} versions of the original LG sequence. Our protocol incorporates the advanced LG4 sequence~\cite{Halse13} over nuclei, which exhibits remarkable decoupling rates and enhanced robustness against RF control errors. The LG4 consists on concatenated blocks of four consecutive off-resonant RF drivings, all complying with the LG condition, leading to rotations along four different axes. This is, the rotation axis $P$ alternates among $A,\bar A,\bar B$ and $B$, whose relative positions are illustrated in Fig.~\ref{rotations}~(a). Note that, at each block, nuclear spins undergo two sets of complementary rotations along axis pointing in opposite directions ($A, \bar A$ and $B, \bar B$). To further illustrate the journey of the magnetization during an LG4 block we include an animation~\cite{supp1}.

The nuclear spin Hamiltonian during each individual rotation of the LG4 sequence reads (see Supplementary Information~\cite{Supp}) 
\begin{equation}\label{OnerotationH}
H = \sum_{i=1}^N\left(\frac{\pm\delta_i}{\sqrt{3}} + \bar{\Omega}\right) I^i_P,
\end{equation}
where $\delta_i$ is the {\it target nuclear shift} of the $i$th spin (its sign depends on the direction of the rotation, positive value $``+\delta_i"$ is assigned to rotations along $A$ and $B$, and the negative value $``-\delta_i"$ to rotation along $\bar A$ and $\bar B$), $\bar{\Omega}=\sqrt{\Delta^2+\Omega^2}$ is the effective Rabi frequency, and the spin operator $I^i_P$ takes one of the following forms
\begin{eqnarray}\label{rotaxes}
I^i_A  &=&  \left(\Omega I^i_x\sin{\alpha}+\Omega I^i_y\cos{\alpha}+ \Delta I^i_z\right)/\bar{\Omega}, \nonumber\\
I^i_{\bar{A}} &=&   - I^i_A,\nonumber\\
I^i_B  &=&  \left(-\Omega I^i_x\sin{\alpha}+\Omega I^i_y\cos{\alpha}+ \Delta I^i_z \right)/\bar{\Omega},\nonumber\\
I^i_{\bar{B}} &=& - I^i_B.
\end{eqnarray} 
According to the LG4 scheme~\cite{Halse13}, the phase of the driving is set to $\alpha=55^\circ$ to minimize the line-width of the resonances. 

Note that in Eq.~(\ref{OnerotationH}) we assume that the internuclear interaction Hamiltonian $H_{\rm nn}=\sum_{i>j}^N \frac{\mu_0\gamma^2_n \hbar}{4\pi r_{i,j}^3} \bigg[\vec{I}_i \cdot \vec{I}_j - 3 (\vec{I}_j  \cdot \hat{r}_{i,k})  (\vec{I}_j \cdot \hat{r}_{i,j})\bigg]$ can be neglected due to the introduced decoupling sequence. This assumption simplifies the subsequent analysis. However, $H_{\rm nn}$ will be taken into account in the numerical model in the results section.

In the remainder of this section, we analyze the signal emitted by the sample subjected to the RF decoupling fields and develop analytical expressions for the target energy shifts.

The magnetic field that originates from the sample during the nuclear spin rotation produced by each RF field of the LG4 follows the general form
\begin{equation}\label{eq:Signal}
s(t) = \Gamma\cos{(\bar\Omega t+\phi)}+b.
\end{equation}
Hereafter, we often refer to $s(t)$ as the signal, as it constitutes the target field for the NV ensemble sensor. In fact, its amplitude $\Gamma$, phase $\phi$, and static bias $b$ depend on the configuration of the nuclear spin ensemble and thereby on the $\delta_i$ energy shifts (see~\cite{Supp}), so detecting and properly reading $s(t)$ enables to unravel the desired information.

Consequently, the LG4 meets a twofold goal. Namely: (i) It results in a nuclear spin dynamics with minimal effect from the dipole-dipole interaction (see~\cite{Supp}for the full derivation of the Hamiltonian in Eq. \eqref{OnerotationH}), enabling the identification of the weaker but interesting $\delta_i$ shifts. (ii) It induces a tunable rotation speed in the sample ($\propto \bar{\Omega}$, see Eq.~(\ref{eq:Signal})), facilitating the interaction between nuclear spins and the NV ensemble sensor even at high external magnetic fields. Regarding point (ii), it is important to note that without using RF drivings on the sample, standard techniques based on imprinting in the NVs a rotation speed comparable to the nuclear Larmor frequency would necessitate the application of unrealistic MW fields. For context, in a magnetic field of approximately 2.35 Tesla, hydrogen spins rotate at a speed of $(2\pi)\times 100$~MHz, producing a signal hardly trackable by an NV ensemble sensor operating with conventional methods~\cite{Glenn18, Bucher20, Arunkumar21}.

Now, we examine the effects of RF decoupling fields in greater detail. Each RF driving (leading to the rotations along $A, \bar A$, $B, \bar B$) is applied for an interval $T = 1/\bar\Omega$. Consequently, the total signal emitted by the sample is a composite of distinct sinusoidal functions, condensed in Eq.~\eqref{eq:Signal}, each persisting for a duration $T$. Figure~\ref{rotations} (b) presents an illustrative example of $s(t)$ by showing the rotation of a single magnetization vector (associated with a specific $\delta_i$) around axes $A$, $\bar{A}$, $B$  and $\bar B$. 

Interestingly, with this RF control, the nuclear spins governed by Eq.~\eqref{OnerotationH} would perform a complete turn at each RF driving, constantly returning to their initial configuration if it were not for the $\delta_i$ shifts. These shifts slightly alter the nuclear spin state (i.e., the sample magnetization), thus imprinting a slower motion within the sample.
More specifically, the sample magnetization at the end of each LG4 block is determined by a set of energy shifts $\delta^*_j$ (distinct from $\delta_i$) according to the effective Hamiltonian:
\begin{equation}\label{eq: Heff}
H_{\rm eff} = \sum_i \delta^*_i I^i_C,
\end{equation}
where $I^i_C $ is a spin operator along an axis $C$ that bisects $A$ and $B$, see  Fig.~\ref{rotations}~(a), while 
\begin{equation}\label{eq:newshifts}
    \delta_i^* = \delta_i \frac{\sqrt{1 + 2 \cos^2{\alpha}}}{3}.
\end{equation}

In summary, this section demonstrates that each LG4 block alters the sample magnetization $\vec{M}$ through rotations along the $C$ axis, as depicted in Fig.~\ref{rotations} (a). An animation of the magnetization precession around the $C$ axis (leading to the yellow rotation plane Fig.~\ref{rotations} (a)) is available in~\cite{supp2}. Moreover, we elucidate the mechanism governing the evolution of $\vec{M}$ through the effective Hamiltonian outlined in Eq.$\eqref{eq: Heff}$, while Eq.~\eqref{eq:newshifts} establishes   analytical expressions connecting the rates of the effective rotations, $\delta^*_i$, with the target nuclear shifts $\delta_i$.

In the next section we outline the protocol to monitor this effective precessions with the NV ensemble sensor and extract the desired $\delta_i$ energies from its recordings. 

\subsection*{Harvesting nuclear spin parameters with the NV ensemble}

\subsubsection*{Geometrical interpretation of the phase accumulation}
\begin{figure}[t!]
\includegraphics[width= .9 \linewidth]{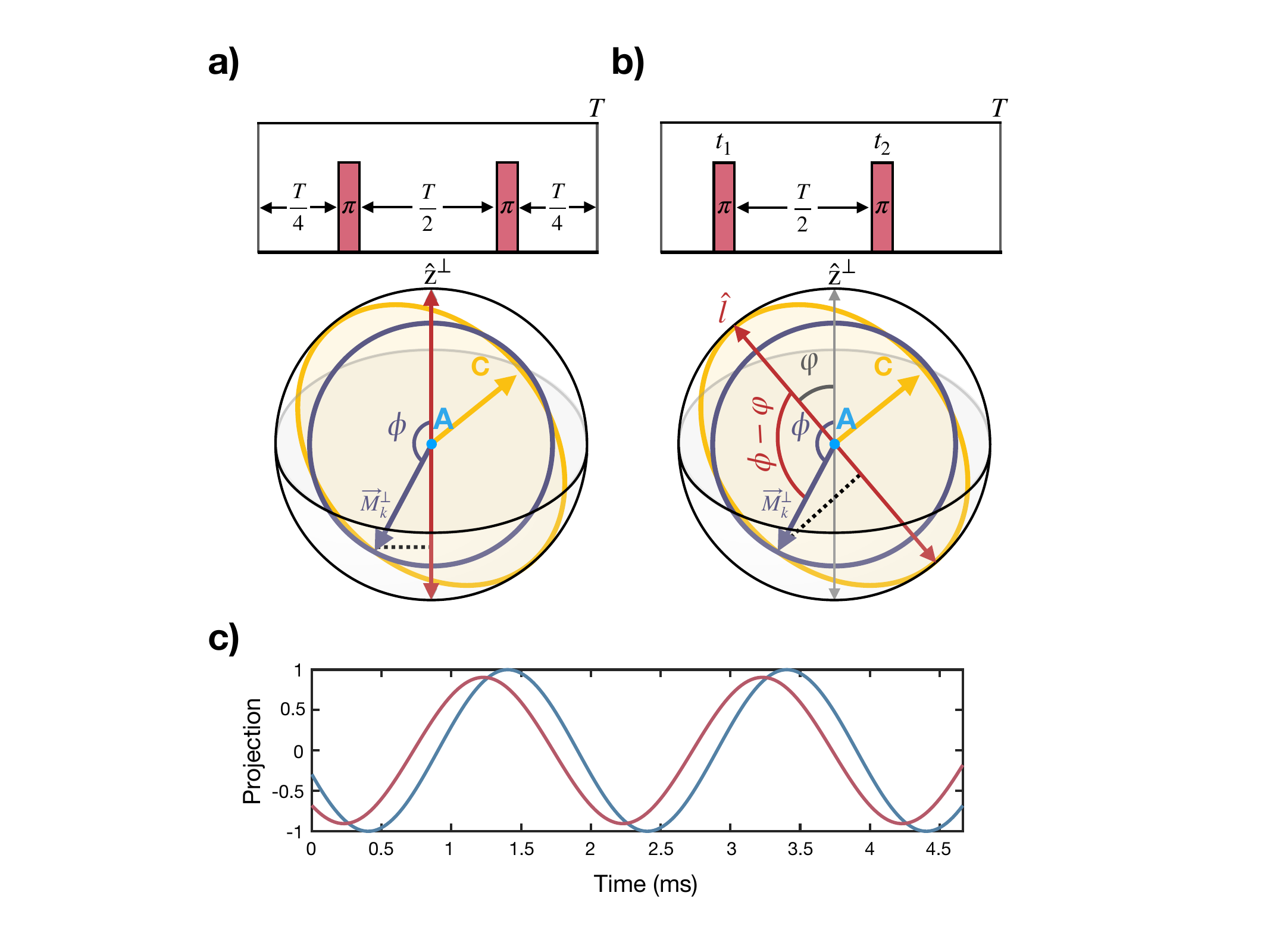}
\centering
\caption{\label{control} (a) Illustration of a CPMG pulse block (upper panel) and its geometric interpretation (lower panel). This panel shows a projection of the sphere in Fig.~\ref{rotations} (b)  viewed in a direction parallel to axis $A$, as represented with the eye symbol. This view facilitates the representation of the projections onto the plane perpendicular to A of (i) The magnetization vector, denoted as $\vec{M}_k^\perp$, and (ii) The $\hat z$ axis, referred to as $\hat z^{\perp}$. In addition, it shows the trajectory followed by a magnetization vector during a rotation around $A$ (purple circle) and after successive LG4 blocks (yellow ellipse). (b) Upper panel, pulse block of our tailored sequence where the initial $\pi$ pulse is delivered at a time $t_1$. With this control, the phase accumulation of the NV is proportional to the projection of  $\vec{M}_k^\perp$ onto $\hat l$ (shown in red), an axis tilted away from $\hat z^\perp$ and aligned with the major axis of the yellow ellipse for optimal contrast. (c) Evolution of the projection of $\vec{M}_k^\perp$ onto axis $\hat z^\perp$ (red) and onto axis $\hat l$ (blue). The amplitude of the projection onto $\hat l$, resulting from the sequence in panel (b), reaches the maximum value of 1. As the phase accumulated by the NV is directly proportional to this projection, the timing of the pulses in (b) ensures the maximum phase accumulation amplitude.}
\end{figure}
The target magnetic field over the NV ensemble sensor is a concatenation of the sinusoidal signals in Eq.~\eqref{eq:Signal} (see lower panel in Fig.~\ref{rotations}(b)). 
A particular RF field at the $k^\text{th}$ LG4 block (note that, the accumulative character of the rotations imposed by Eq.~\eqref{eq: Heff} make it crucial to identify the number of the block from now on), produces a nuclear spin rotation around a certain axis ($A$, $\bar A$, $B$ or $\bar B$) where the amplitude $\Gamma_k$ of the resulting signal $s_k(t) = \Gamma_k\cos{(\bar\Omega t+\phi_k)}+b_k$ is directly proportional to $\vec{M}_k^\perp$ (i.e., to the magnetization component which is orthogonal to the rotation axis --$A$, $\bar A$, $B$ or $\bar B$-- at the start of each RF driving), and the phase $\phi_k$ corresponds to the angle between $\vec{M}_k^\perp$ and $\hat z^\perp$. The latter is the component of $\hat z$ that lies on the plane perpendicular to the rotation axis. See the lower panel in Fig.~\ref{control} (a) and~\cite{Supp} for more details.

We now use this geometric description to analyze the phase accumulated by each NV in the ensemble sensor when subjected to a generic pulse sequence. For this analysis, we choose a standard Carr-Purcell-Meiboom-Gill (CPMG) sequence~\cite{Purcell46, Meiboom58}. In order to keep the discussion accessible, we focus on the signal produced by the nuclear spins rotating around $A$ and limit ourselves to an scenario involving a single effective energy shift, $\delta_i^*$. Note, however, that the following results and the consequent conclusions are valid for signals produced by nuclear spins rotating around axes $B$, $\bar A$ and $\bar B$ and in situations involving multiple shifts.

The phase accumulated by an NV center interacting with the signal $s_k(t)$ and subjected to the CPMG sequence reads (see~\cite{Supp})
\begin{equation}\label{eq: phase}
\Phi = \frac{4 |\gamma_e|}{\bar \Omega}\Gamma_k \cos{\left(\phi_k\right)}.
\end{equation}
Hence, the phase accumulated by each NV is proportional to the projection of $\vec{M}_k^\perp$ onto  $\hat z^\perp$, or, in other words, to the quantity $\Gamma_k \cos{\left(\phi_k\right)}$. The lower panel of Figure~\ref{control} (a)  provides a clarifying (probably most needed) graphic explanation.

With this description in mind we can summarize the phase acquisition stage as follows: The response of the NV centers to the signal emitted by the sample is determined by the initial sample magnetization. As the protocol advances, the magnetization vector precesses around $C$ with an angular velocity $\delta^*_i$ as described by Eq.~(\ref{eq: Heff}). In the orthogonal plane with respect to $A$, this precession translates into an elliptical motion of the vector $\vec{M}_k^\perp$, shown as a yellow ellipse in Fig.~\ref{control} (a). Thus, the projection of $\vec{M}_k^\perp$ onto the $\hat z^\perp$ axis, and consequently the phase accumulated by the NV in successive blocks of the LG4, follow a sinusoidal function with frequency $\delta^*_i$. The resulting expected value of the $\sigma_z$ operator of each NV in the ensemble at the $k^\text{th}$ LG4 block (after applying a final $\pi/2$ pulse to transform accumulated phase into populations), generalized to every $\delta_i^*$, reads: 
\begin{equation}\label{eq:expectedsigmaz}
\langle \sigma_z\rangle_k \approx 3 D_0 \sum_i \rho_i\cos{\left(\frac{4\delta_i^* k}{\bar{\Omega}} + \nu_0\right)},
\end{equation}
 where $\rho_i$ is the spin density of the $i$th nucleus, $D_0$ is detemined by the pulse sequence and $\nu_0$ depends on both the pulse sequence and the initial nuclear state. A formal derivation of \eqref{eq:expectedsigmaz} as well as further details can be found in~\cite{Supp}. The $0$ subindex in Eq.~(\ref{eq:expectedsigmaz}) indicate that all parameters correspond to the reference CPMG sequence (note that, in the next section we derive an improved sequence). Thus, the NV response  enclosed in Eq.~(\ref{eq:expectedsigmaz}) consists on a sum of sinusoidal functions that encode the different $\delta_i^*$ which can be then extracted via standard Fourier transform. Finally, $\delta_i$ targets can be obtained via a direct application of Eq.~(\ref{eq:newshifts}).  

\begin{figure*}[t]
\centering
\includegraphics[width=0.85\linewidth]{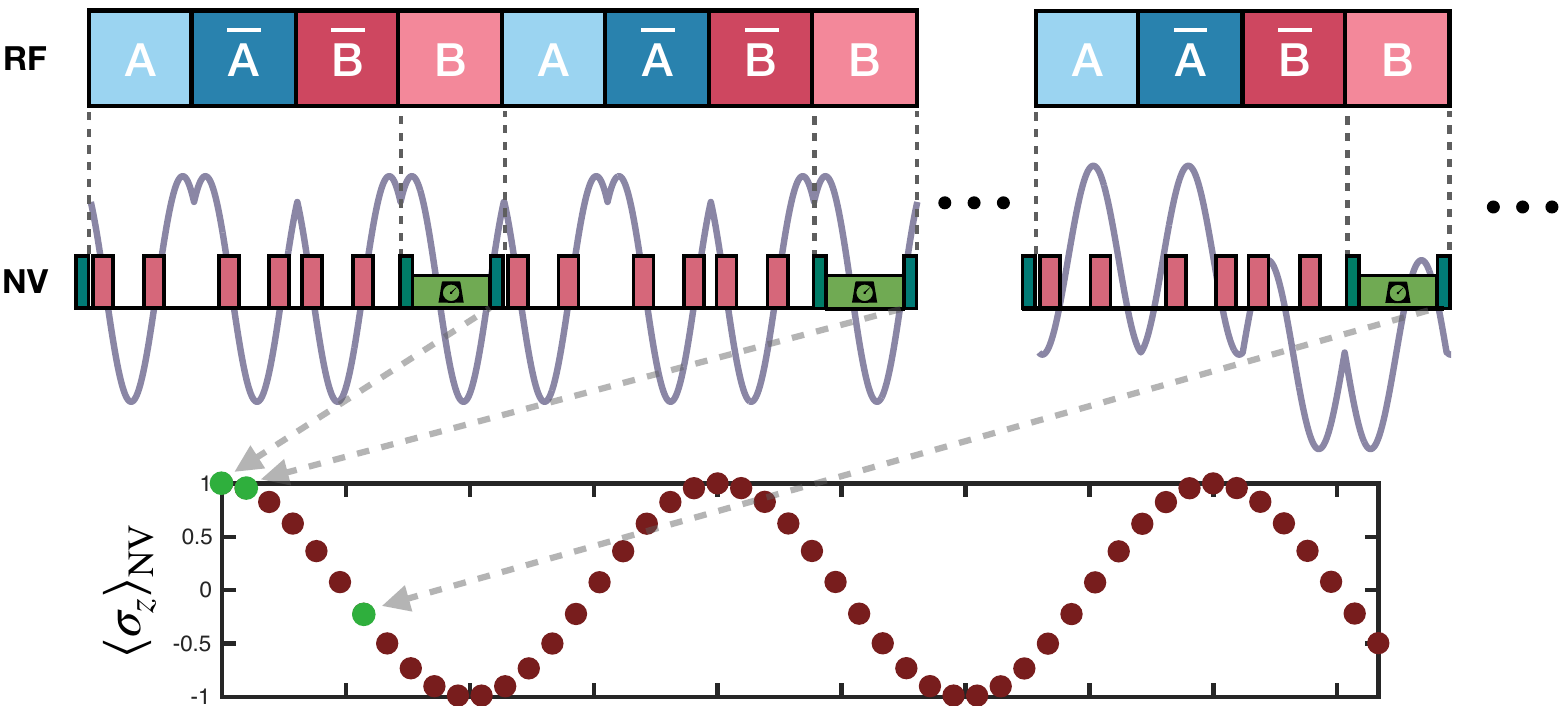}
\caption{\label{scheme} (Top) General layout of our protocol containing the RF control over nuclear spins and the MW pulse sequence on the NV ensemble. The magnetic field emitted by the nuclei as a consequence of the RF rotations is depicted in light purple. As time progresses, this field changes its shape, which changes the phase accumulated by the NV, while MW pulses keep always the same structure. (Bottom) Evolution of the expected results for the measurement over the NVs as the experiments progresses, showing a sinusoidal pattern with frequency $\delta^*$. In the presence of additional shifts, the measurement outcomes evolve as a sum of sinusoidal components with corresponding frequencies $\delta^*_i$, which can be extracted through Fourier transform analysis.}
\end{figure*}
\subsubsection*{Sensing MW pulse sequence}

With the geometrical understating developed in the previous section, now we present a tailored MW sequence   to optimally detect the target $\delta_i$ shifts. This sequence retains a CPMG-like structure composed by two $\pi$ pulses spaced by $T/2$ to mitigates noise effects, $T$ being the CPMG block length. Nonetheless, we adjust the timing of the pulses, see  Fig.~\ref{control} (b), in particular the time at which the first pulse is applied ($t_1$ hereafter). Consequently, the phase accumulated by each NV in the ensemble sensor (recall that we are focusing on the signal produced by the nuclear spins rotating around $A$) reads  
\begin{equation}\label{eq: optimal}
\Phi = \frac{4\Gamma_k|\gamma_e|}{\bar{\Omega}}\cos{\left(\phi_k-\varphi\right)}.
\end{equation}
In our geometrical framework, adjusting the timing of the pulses results in an accumulated phase proportional to the projection of $\vec{M}_k^\perp$ onto an axis $\hat l$, which is tilted at an angle $\varphi = \frac{\pi}{2} - \bar{\Omega} t_1$ relative to $\hat{z}^\perp$ (see Fig.~\ref{control} (b)).

The ability to pivot the axis in which $\vec M_k^\perp$ gets projected (note this can be done by selecting distinct values for $t_1$ since $\varphi = \frac{\pi}{2} - \bar{\Omega} t_1$) allows us to design a pulse sequence that maximizes contrast in the recorded spectra. From block to block, $\vec{M}_k^\perp$ evolves following an ellipse, therefore, we design the pulse sequence so that the phase accumulated by each NV is proportional to the projection of $\vec{M}_k^\perp$ into the major axis of the elipse. By doing so, the projecting axis and the direction that contains the extreme points of the elliptic path of $\vec{M}_k^\perp$ match, thereby yielding the maximum amplitude in the oscillation of $\Phi$ in successive blocks, see Fig.~\ref{control} (b). In particular, this is achieved by setting $\varphi = \arccos{\frac{\sqrt{3}\cos{\alpha}}{\sqrt{2 + \cos{2\alpha}}}}$, which determines the timing of the pulses as $t_{1,A}\approx 0.14\, T$ for optimal detection of the signal produced by the nuclear spins rotating around $A$. Repeating the same analysis for the signals produced by nuclear spin rotations around $\bar A$ and $\bar B$ we find $t_{1, {\bar A}} = \frac{T}{2}-t_{1, A}$ and $t_{1, {\bar B}} = t_{1, A}$. 

Summing up, our tailored MW pulse sequence is separated in blocks. Each block contains two $\pi$ pulses specifically timed to optimally detect the signal produced by the corresponding RF field. To maintain synchrony between the two control channels (MW and RF) and to avoid turning off the nuclear decoupling field, the sensor is measured and reinitialized while the RF field is on. Figure~\ref{scheme} shows the general layout of our protocol, including the drivings over the sample and the sensor, and showing the evolution of the expected outcomes in successive measurements, which read 
\begin{equation}
\langle \sigma_z\rangle_k \approx 3 D_{\rm opt} \sum_i \rho_i\cos{\left(\frac{4\delta_i^* k}{\bar{\Omega}} + \nu_{\rm opt}\right)},
\end{equation}
where $D_{\rm opt}\approx 1.1 D_0$ (i.e., with the tailored MW sequence the contrast increases a $10\%$) and $\nu_{\rm opt} = 0$ which corresponds to a initial sample magnetization oriented along the axis perpendicular to the $A$ and $B$ axes, achieved by a RF pulse that triggers the protocol. Finally, we access the effective frequencies $\delta_i^*$ by Fourier transforming the recorded data and obtain the target $\delta_i$ shifts from Eq. \eqref{eq:newshifts}.

\begin{figure*}[t]
\centering
\includegraphics[width= 0.9 \linewidth]{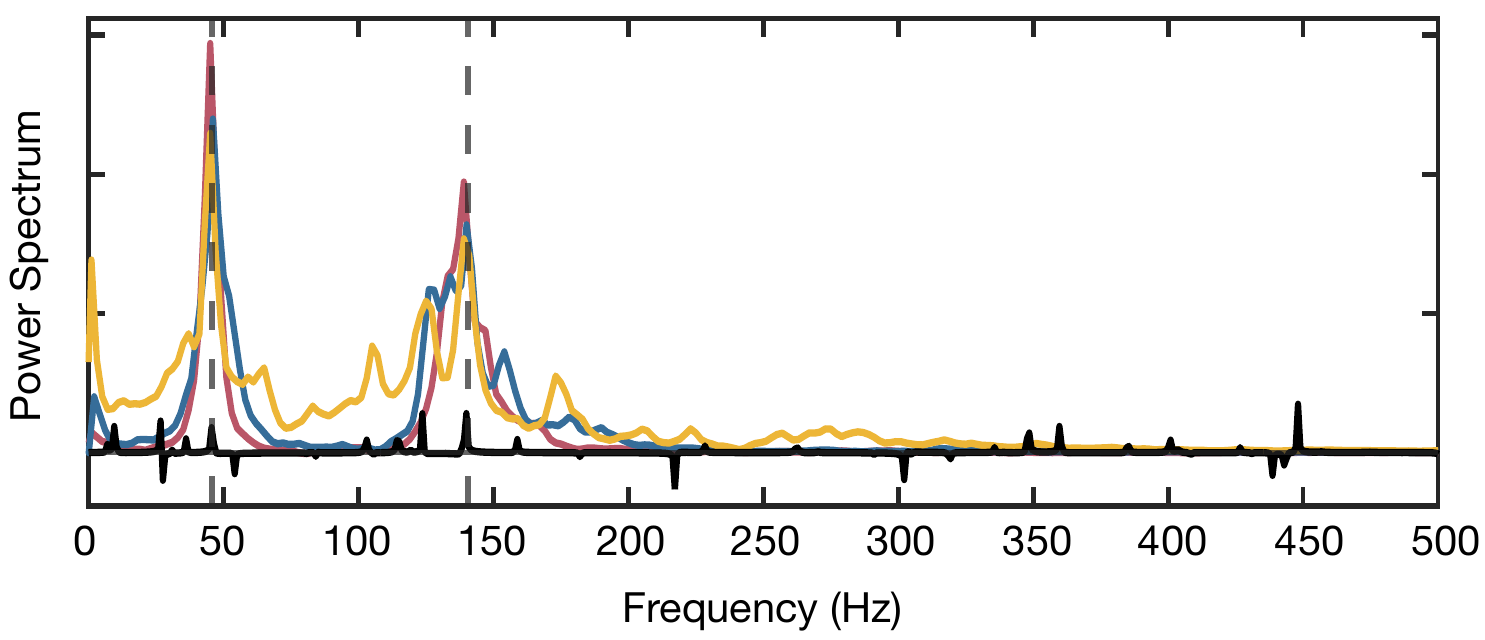}
\caption{\label{results} Spectra obtained from three simulations with RF Rabi frequencies of $(2\pi)\times$100 kHz (yellow line), $(2\pi)\times$150 kHz (blue line), and $(2\pi)\times$200 kHz (red line). Vertical dashed gray lines indicate the expected resonances, i.e. the two $\delta^*_i$, of approximately 140 Hz and 45 Hz. For comparison, a spectrum obtained with AERIS is depicted (black line), using an RF Rabi frequency of $(2\pi)\times$150 KHz. In all simulations, the nuclear sample starts in a thermal state corresponding to a temperature of T=300 K in an external magnetic field of 2.1 T. The sample evolves for a total time of 0.5 s.}
\end{figure*}

\section*{Results}\label{ResultsSec}

We test our protocol by simulating its implementation to detect the chemical shifts of hydrogen nuclear spins in ethanol molecules (${\rm C_2H_6O}$) at high external magnetic field. Although ethanol typically exists as a liquid, we employ its solid configuration as an example of an ordered sample with strong homonuclear dipole-dipole couplings. In particular, ethanol molecules exhibit dipolar couplings of up to 17 kHz (the proton attached to the oxygen shows limited dipole interaction with the rest of the system so we exclude it from the simulations). High external fields improve nuclear magnetic resonance procedures, not only because it yields higher polarization rates, but also because it enhances the weaker, and thus harder to detect, energy shifts. In our simulations, we consider an external field of $B_0 = 2.1$ T, with chemical shifts of 3.66 ppm for three of the protons and 1.19 ppm for the other two, corresponding to frequency shifts of approximately 327 Hz and 106 Hz, respectively.

Our numerical simulation unfolds in two phases. First we find the target magnetic signal by simulating the evolution of the nuclear spin sample subjected to the LG4  decoupling sequence. The approximate Hamiltonian in 
Eq.~(\ref{OnerotationH}) facilitates a deeper understanding of the nuclear dynamics and in particular the development of the geometrical interpretation that has set the ground for the design of our protocol. Our simulations, however, make use of the exact nuclear spin Hamiltonian, which reads
\begin{equation}
\begin{aligned}
H(t) &= \sum_{i = 1}^{N = 5} \left\{-\gamma_h \delta_i B_0 I_z^i  + \left[\Omega+\eta(t)\right] I_\phi^i + \Delta I_z^i \right\}+\\
 &+ \sum_{i > {j = 1}}^{N = 5}-\frac{\mu_0 \hbar \gamma_n^2}{8 \pi |r^{ij}|^3}\left(3 r^{ij}_z-1\right)\left[3 I_z^i I_z^j - \vec{I}_i\cdot\vec{I}_j\right],
\end{aligned}
\end{equation}
where $\delta_i$ is the target nuclear shift of the $i$th hydrogen atom and $\eta(t)$ is the driving noise modeled with an Ornstein-Uhlenbeck process with a $1$ ms correlation time and 0.24\% amplitude. 

We assume the sample starting in a completely mixed state. A triggering RF pulse sets the desired initial state $\rho(0)$, a thermal state oriented along the axis perpendicular to $A$ and $B$. From there, the evolution of the nuclear density matrix is simulated via the master equation
\begin{equation}
\dot{\rho} = -i\left[H, \rho\right] + \frac{1}{2 T^*_2}\sum_{j = 1}^{N = 5}\left(4 {I}^j_z \rho {I}^j_z-\rho\right),
\end{equation}
where $T^*_2 = 0.2$ s is chosen to approximate the intrinsic coherence time in the absence of dipolar couplings (which are explicitly included in the Hamiltonian). This equation leads to a signal computed as
\begin{equation}
s(t) = \frac{\gamma_h \hbar \mu_0 \sigma_h g}{4\pi} \operatorname{Tr}\left[\rho(t)\bar{I_z}\right],
\end{equation}
where $\sigma_h = 5.2\times 10^{28} \mathrm{m^{-3}}$ is the number density of hydrogen spins for ethanol, and $g \approx~4.1$ is a geometric factor that relates the sample magnetization and the magnetic field in the NV location, see \cite{Glenn18, Munuera23} for further details.

In the second phase, we simulate the evolution of the NV ensemble interacting with $s(t)$ and subjected to the sensing MW pulse sequence of our protocol. The Hamiltonian that governs the dynamics of each NV centers reads
\begin{equation}
H = \gamma_e s(t)\frac{\sigma_z}{2}+C(t)\frac{\sigma_z}{2}+\frac{\Omega_{\rm NV}(t) \sigma_\phi}{2},
\end{equation}
with $\Omega_{\text{NV}}(t)$  the control Rabi frequency, and $C(t)\frac{\sigma_z}{2}$ describing potential RF-induced crosstalk on the NVs. Following the protocol devised in this work, they interact with the signals produced by the nuclear spin rotations around axis $A$, $\bar A$ and $\bar B$ and accumulate a phase determined by the state of the sample magnetization. During the time that corresponds to the delivery of the $B$ RF field over the sample, we transform the accumulated phase into a population difference with a $\pi/2$ pulse and simulate the measurement, after which the sensor is reinitialized and the protocol repeats. Finally, a Fourier transform of the measurements provides the spectra displayed in Fig. \ref{results}, which proves the ability of our protocol to access the chemical shifts of the molecule.

Figure~\ref{results} shows three different experiments with increasing RF intensities. As expected, stronger RF drivings lead to clearer spectra, less distorted by spurious peaks which arise due to incomplete averaging of dipolar couplings. In particular we simulate RF drivings with Rabi frequencies of $(2\pi)\times$100 kHz, $(2\pi)\times$150 kHz, and $(2\pi)\times$200 kHz, all attainable values by state of the art antennas \cite{Herb20, Yudilevich23}, and set the Rabi frequency of the MW control at 20 MHz in all cases. Note that when the RF Rabi frequency exceeds the 17 KHz dipolar coupling strength by an order of magnitude, the resulting spectra become significantly cleaner and more resolved. As intended, our method leads to resonance peaks centred in $\delta_i^*$ from which one can extract the target nuclear shifts $\delta_i$ using Eq. \eqref{eq:newshifts}. For comparison, we include the results of a fourth simulation using AERIS~\cite{Munuera23}, which does not incorporate any dipolar coupling suppression technique. In this case, the obtained spectra (black-curve) is distorted as a consequence of the strong nuclear dipolar couplings. 

In summary, we have demonstrated that our protocol effectively identifies the target energy shifts $\delta_i$  with strong nuclear dipole-dipole interactions at high external magnetic fields.

\section*{Conclusions}
We have designed a protocol that utilizes LG4 sequences and a tailored NV pulse train to identify chemical shifts in the presence of strong dipole-dipole interactions. The RF field serves two key purposes: (i) decoupling nuclear spins and (ii) generating a nuclear signal oscillating at a moderate frequency that can be measured by the NVs, allowing the protocol to operate at high magnetic fields. By incorporating a tailored MW sequence on the NV for signal detection, we achieve effective retrieval of chemical shifts. Finally, the accuracy of our method is ultimately limited by the nuclear sample decoherence, thus surpassing the limitations imposed by NVs dephasing and thermalization. Our findings pave the way for the advancement of microscale NMR techniques and broaden their application in diverse fields, such as materials science, chemistry, and biology.

\section*{Acknowledgements}
C.M.-J. acknowledges the predoctoral MICINN grant PRE2019-088519. J. C. acknowledges the Ram\'on y Cajal (RYC2018-025197-I) research fellowship. Authors acknowledge the Quench project that has received funding from the European Union's Horizon Europe -- The EU Research and Innovation Programme under grant agreement No 101135742, the Spanish Government via the Nanoscale NMR and complex systems project PID2021-126694NB-C21, and the Basque Government grant IT1470-22. A.T. acknowledges support from the Horizon Europe project QCircle 101059999 (Teaming for Excellence).

\section*{Data availability}
The code used to produce the results in this study is available at https://github.com/carlosmunueraj/NV-LG4-DD-samples.

\newpage
\onecolumn
\appendix

\section{Nuclear spin dynamics under RF drivings~\label{app1}}

Under an RF field (that will rotate nuclei around the $A , \bar{A}, B,$ or $\bar{B}$ axes) the nuclear spin Hamiltonian including dipole-dipole terms among nuclear spins reads
\begin{eqnarray}\label{HS}
H =  \sum_i \left[ \gamma_n B_z I^i_z +  \delta_j I^i_z + 2 \Omega I^i_x \sin{\left(\omega_d t - \alpha\right)} \right] 
+\sum_{i>j} \frac{\mu_0\gamma^2_n \hbar}{4\pi r_{i,j}^3} \bigg[\vec{I}_i \cdot \vec{I}_j - 3 (\vec{I}_i  \cdot \hat{r}_{i,j})  (\vec{I}_j \cdot \hat{r}_{i,j})\bigg],
\end{eqnarray}
where $r_{i,j}$ is the distance between each pair of nuclei ($\hat{r}_{i,j}$ its a unitary vector such that $\vec{r}_{i,j} = r_{i,j}\hat{r}_{i,j}$), $\gamma_N$ is the nuclear gyromagnetic ratio, $\mu_0$ is the vacuum permeability, $\vec I_i ={\left(I^i_x,I^i_y,I^i_z\right)}$ are the nuclear spin operators for the $i$th spin, $\Omega$ and $\omega_d$ are the Rabi and carrier frequencies of the radio-frequency (RF), $\alpha$ is a tunable phase of the RF, and $\delta_i$ represents the deviation (i.e. the energy shift) of the $i$th nuclear spin from the Larmor precession rate $\omega_L = \gamma_n B_z $ owing to its particular magnetic environment. The accurate determination of $\delta_j$ is the target of the sensing protocol introduced here. 

In a rotating frame w.r.t. $ (\omega_L - \Delta) \sum_i I_z^i$, and under the secular approximation of the dipole-dipole term that eliminates fast rotating terms by invoking the rotating wave approximation, Eq.~(\ref{HS}) simplifies to

\begin{eqnarray}\label{RFin}
H = \sum_i \left[ (\Delta + \delta_i) I_z^i + \Omega I^i_\alpha \right]
+ \sum_{i>j} \frac{\mu_0\gamma^2_n \hbar}{4\pi r_{i,j}^3} \left[1-3\left(r_z^{i,j}\right)^2\right] \left[I^i_z I^j_z -\frac{1}{2} (I^i_\alpha I^j_\alpha + I^i_{\alpha^\perp} I^j_{\alpha^\perp})\right].
\end{eqnarray}
where $I^i_\alpha = \left(I^i_x \sin\alpha + I^i_y \cos\alpha\right)$, and ${I^i_{\alpha}}^\perp = \left(I^i_x \cos\alpha - I^i_y \sin\alpha\right)$.

Introducing a new spin basis, defined by rotating the original axes around $\alpha^\perp$, 
\begin{eqnarray}\label{change}
&&I_P^j = \cos{(\theta)}I_z^j + \sin{(\theta)}I_\alpha^j,\nonumber\\
&&I_Q^j = \cos{(\theta)}I_\alpha^j - \sin{(\theta)}I_z^j,\nonumber\\
&&I_{Q^{\perp}}^j = I_{\alpha^{\perp}}^j,
\end{eqnarray}
and defining the rotation angle through  $\cos{\theta}=\frac{\Delta}{\sqrt{\Omega^2 + \Delta^2}}$, and $\sin{\theta} = \frac{\Omega}{\sqrt{\Omega^2 + \Delta^2}}$, allows to rewrite the sample Hamiltonian as

\begin{equation}
\begin{aligned}\label{hlg}
&H = \bar \Omega \sum_i I_P^i + \sum_i \delta_i[\cos{(\theta)}I_P^i -  \sin{(\theta)}I_Q^i]
+ \sum_{i>j} \frac{\mu_0\gamma^2_n \hbar}{4\pi r_{i,j}^3} \left[1-3(r_{i,j}^z)^2\right] \Bigg\{\cos^2{(\theta)}I_P^i I_P^j  + \sin^2{(\theta)}I_Q^i I_Q^j\\
& - \cos(\theta) \sin(\theta)(I_P^i I_Q^j + I_Q^i I_P^j )-\frac{1}{2}\bigg[\cos^2{(\theta)}I_Q^i I_Q^j
 + \sin^2{(\theta)}I_P^i I_P^j + \cos(\theta) \sin(\theta)(I_P^i I_Q^j + I_Q^i I_P^j )+ I_{Q^{\perp}}^i I_{Q^{\perp}}^j\bigg]\Bigg\},
\end{aligned}
\end{equation}
where the effective rotation rate around $I_P$ reads $\bar \Omega = \sqrt{\Delta^2 + \Omega^2}$. Finally, many terms can be neglected by a secular approximation with respect to $\bar \Omega \sum_i I_P^i$. The remaining terms in the dipolar interaction disappear {\it magically} when the angle that defines the change of basis in Eq. \eqref{change} satisfies $\cos(\theta) = \pm 1/\sqrt 3$, or, equivalently, when the Lee-Goldburg condition $\Delta = \pm \Omega/\sqrt{2}$ is met, leading to 

\begin{equation}
H = \sum_{i=1}^N\left(\frac{\pm\delta_i}{\sqrt{3}} + \bar{\Omega}\right) I^i_P,
\end{equation}

where $\pm\delta_j/\sqrt{3}$ are the parallel components of the shifts with respect to the effective rotation axis {\it P}, and its sign is the same as the sign of $\Delta$. Note that any combination of $\Omega$ and $\Delta$ that complies with the Lee-Goldburg condition produces the described decoupling effect. In particular, for a given intensity of the RF field, this can be detuned from the top and from the bottom with respect to the Larmor. Moreover, the previous derivation is valid for any phase $\alpha$ of the RF field. This freedom has been exploited to develop more elaborated control schemes that concatenate various RF fields, such as the LG4 sequence implemented in our protocol.

In the LG4 sequence, each driving axis is applied during a time $T=2\pi/{\bar\Omega}$ following the order $A$, $\bar A$, $\bar B$, and $B$ (see main text). In order to obtain the effective dynamics of a full LG4 block, we write the explicit propagator
\begin{equation}
U_{\rm LG4} = U_{B}U_{\bar B}U_{\bar A}U_A =
e^{-i\sum_{i=1}^N\left(\frac{\delta_i}{\sqrt{3}} + \bar{\Omega}\right) I^i_B T}e^{-i\sum_{i=1}^N\left(-\frac{\delta_i}{\sqrt{3}} + \bar{\Omega}\right) I^i_{\bar B}T}e^{-i\sum_{i=1}^N\left(-\frac{\delta_i}{\sqrt{3}} + \bar{\Omega}\right) I^i_{\bar A}T}e^{-i\sum_{i=1}^N\left(\frac{\delta_i}{\sqrt{3}} + \bar{\Omega}\right) I^i_A T}.
\end{equation}
In every propagator, we can do the following change
\begin{equation}
e^{-i\sum_{i=1}^N\left(\pm\frac{\delta_i}{\sqrt{3}} + \bar{\Omega}\right) I^i_P T}=e^{-i\sum_{i=1}^N\pm\frac{\delta_i}{\sqrt{3}} I^i_P T}e^{-i\sum_{i=1}^N\bar \Omega I^i_P T}=e^{-i\sum_{i=1}^N\pm\frac{\delta_i}{\sqrt{3}} I^i_P T},
\end{equation}
where we used that $e^{-i\sum_{i=1}^N\bar \Omega I^i_P T}=e^{-i\sum_{i=1}^N 2\pi I^i_P}=\mathbb{I}$. Assuming that $\bar\Omega>>\pm\frac{\delta_i}{\sqrt{3}}$, we can Trotterize the LG4 propagator to obtain
\begin{equation}
e^{-i\sum_{i=1}^N\left(\frac{\delta_i}{\sqrt{3}} I^i_B-\frac{\delta_i}{\sqrt{3}} I^i_{\bar B}-\frac{\delta_i}{\sqrt{3}} I^i_{\bar A}+\frac{\delta_i}{\sqrt{3}} I^i_A\right) T}.
\end{equation}
Finally, substituting the expression for each axis operator of Eq.~(2) in the main text, we obtain the propagator
\begin{equation}
e^{-i\sum_{i=1}^N\left[\frac{\delta_i}{\sqrt{3}\bar \Omega} \left(\Omega I^i_y\cos\alpha+\Delta I^i_z\right)\right]4T}.
\end{equation}
From this expression, we reach the final effective Hamiltonian after rearranging the terms
\begin{equation}
H_{\rm eff} = \sum_i \delta^*_i I^i_C,
\end{equation}
where $I^i_C =  \frac{\sqrt{2}I^i_y\cos{\alpha} + I^i_z}{\sqrt{2\cos^2{\alpha} + 1}}$ and $\delta_i^* = \delta_i \frac{\sqrt{1 + 2 \cos^2{\alpha}}}{3}$.

\section{Accumulated Phase\label{app: accumulated}}
As stated in the main text, the signals received by the NV adhere to a general form Eq.~(3) in the main text. When a two pulse CPMG sequence is applied on the NV sensor (see Fig.~(1) in the main text) with $\pi$ pulses applied at times $t_1$ and $t_2$, the phase accumulated by the NV at stage $k$ is:

\begin{equation}
\Phi_k=\int_0^{t_1}\left[|\gamma_e|\Gamma_k\cos{\left(\bar\Omega t + \phi_k\right)}+b_k\right]dt-\int_{t_1}^{t_2}\left[|\gamma_e|\Gamma_k\cos{\left(\bar\Omega t + \phi_k\right)}+b_k\right]dt+\int_{t_2}^{T}\left[|\gamma_e|\Gamma_k\cos{\left(\bar\Omega t + \phi_k\right)}+b_k\right]dt,
\end{equation}

where we choose the separation of both pulses to be $\frac{T}{2}$, which ensures the cancellation of the static $b_0$ term

\begin{equation}\label{NVPhase}
\Phi_k = \frac{2|\gamma_e|\Gamma_k}{\bar \Omega}\left[\sin{(\bar \Omega t_1 + \phi_k)} - \sin{(\bar \Omega t_2 + \phi_k)}\right] = \frac{4|\gamma_e|\Gamma_k}{\bar\Omega}\cos{\left(\phi_k-\varphi\right)},
\end{equation}
with $\varphi =  \frac{\pi}{2}-\bar\Omega t_1$.

As the sample evolves under the LG4 sequence, the amplitude $\Gamma_k$ and phase $\phi_k$ of the NMR signal evolve, see Fig.~\eqref{control} (a, b). If the signal gets projected about some axis, e.g. $\Gamma_k\cos{\phi_k}$, the variation of this projection is a simple sinusoidal function (see Fig.~\eqref{control} (c)) which is exactly what we need in order to extract the information using a discrete Fourier transform. This result can be understood geometrically, see main text.

Once we choose a projection angle axis, we can compute the adequate timing for the CPMG sequence as
\begin{equation}\label{Atiming}
\varphi = \varphi_{\rm opt} \rightarrow \varphi_{\rm opt} = \frac{\pi}{2}-\bar\Omega t_1 \rightarrow t_1 = \frac{\pi}{2\bar\Omega} - \frac{\varphi_{\rm opt}}{\bar\Omega}.
\end{equation}

For optimal pulse positions, we select the angle matching the major axis of the ellipse. This axis is orthogonal to both $\hat A$ and $\hat C$, i.e., $\left(0, -\frac{1}{\sqrt{2 + \cos{2\alpha}}}, \frac{\sqrt{2}\cos{\alpha}}{\sqrt{2 + \cos{2\alpha}}}\right)$. Then, the angle $\theta_A$ is measured with respect to the orthogonal component of $\hat z$ concerning $\hat A$. This angle is:

\begin{equation}
\varphi_{\rm opt} = \arccos{\frac{\sqrt{3}\cos{\alpha}}{\sqrt{2 + \cos{2\alpha}}}}.
\end{equation}

\section{Analytical expression}\label{app: analytical}
Here we provide details of the derivation of the analytical expression for the expected value of the measurements performed with the NV. Our starting point is the fact that the NV will couple to a signal proportional to the $\hat z$ component of the sample magnetization. 

Focusing on the $k$th driving stage around $\hat A$, we can describe the expected signal as:
\begin{equation}\label{eq: signal_origin}
s \propto \hat M(t)\cdot \hat z = \hat M(t)\left(\hat z^\perp\sin{\theta_{\rm LG}} + \hat A\cos{\theta_{\rm LG}}\right),
\end{equation}
where $\hat M(t)$ is the magnetization vector and the $\hat z$ axis was split in the parallel and perpendicular components with respect to axis $\hat A$, and $\theta_{\rm LG} = \arccos{\frac{1}{\sqrt{3}}}$ is the magic angle. We can describe the time dependency of the magnetization during the driving stage $A$ by employing the Rodrigues' rotation formula as 
\begin{equation}
    \hat{M}(t) = \hat{M}_k\cos{\bar \Omega t}+\left(\hat A\times \hat{M}_k\right)\sin{\bar \Omega t}+\hat A\left(\hat A\cdot \hat{M}_k\right)\left(1-\cos{\bar \Omega t}\right),
\end{equation}
where $\hat M_k$ is the magnetization vector at the beginning of the $k$th sequence. Substituting in Eq. \eqref{eq: signal_origin}, we get 
\begin{equation}
s\propto \left[\hat M_k\cdot \hat z^\perp\cos{\bar \Omega t}+\left(\hat A\times \hat M_k\right)\cdot \hat z^\perp \sin{\bar \Omega t}\right]\sin{\theta_{\rm LG}}+\hat A \cdot \hat M_k \cos{\theta_{\rm LG}}.
\end{equation}
We can now split the magnetization vector into its parallel and perpendicular components with respect to $\hat A$ as $\hat M_k = \left(\vec M_k^{\parallel} + \vec M_k^{\perp}\right)$. With this we reach expression
\begin{equation}
s\propto|\vec M_k^{\perp}|\sin{\theta_{\rm LG}}\cos{\left(\bar \Omega t+\phi\right)}+|\vec M_k^{\parallel}|\cos{\theta_{\rm LG}},
\end{equation}
where $\phi$ is the angle between $\vec M_k^{\perp}$ and $\hat z^\perp$. Notice how this expression exactly matches the shape of Eq.~(3) in the main text.

Substituting in Eq. (8) of the main text, we obtain
\begin{equation}
    \Phi \propto -\frac{4\gamma_e\sin{\theta_{\rm LG}}}{\bar \Omega}|\vec M_k^{\perp}|\cos{\left(\phi-\varphi\right)} = -\frac{4\gamma_e\sin{\theta_{\rm LG}}}{\bar \Omega}\hat M_k\cdot\hat l
\end{equation}
with $\hat l$ a vector perpendicular to $\hat A$ and tilted $\varphi$ with respect to $\hat z^\perp$. 

We can now generalize to all the driving stages by describing the precession motion of the initial magnetization vectors employing Rodrigues' formula once again
\begin{equation}
    \hat{M}_k = \hat{M}_0\cos{\left(\frac{4\delta_i^* k}{\bar{\Omega}}\right)}+\left(\hat C\times  \hat{M}_0\right)\sin{\left(\frac{4\delta_i^* k}{\bar{\Omega}}\right)}+\hat C\left(\hat C\cdot  \hat{M}_0\right)\left[1-\cos{\left(\frac{4\delta_i^* k}{\bar{\Omega}}\right)}\right].
\end{equation}
Starting with an initial magnetization $\hat{M}_0$ in the orthogonal plane with respect to $\hat C$ and an angle $\mu$ with respect to $\hat x$ (which resides in this plane), and including factors for the signal amplitude, we obtain the formula for the accumulated phase
\begin{equation}
    \Phi_k =D_\varphi\rho_i\cos{\left(\frac{4\delta_i^* k}{\bar{\Omega}}+\mu-\beta_\varphi\right)},
\end{equation}
where $D_\varphi = \frac{-\gamma_e\hbar^2\gamma_h^2\mu_0 g \sin{\left(\theta_{\rm LG}\right)}}{8\bar{\Omega}\pi^2 k_B T}\sqrt{\left(\frac{\cos{\varphi}\sin{\alpha}}{\sqrt{3}}-\cos{\alpha}\sin{\varphi}\right)^2+\frac{\left(\sqrt{3}\cos{\alpha}\cos{\varphi}+\sin{\alpha}\sin{\varphi}\right)^2}{2+\cos{\left(2\alpha\right)}}}$, $\rho_i$ is the spin density of the $i$th nucleus, and $\beta_\varphi = \arctan{\frac{3\left(\sqrt{3}\cos{\alpha}\cos{\varphi}+\sin{\alpha}\sin{\varphi}\right)}{\sqrt{2+\cos{(2\alpha)}}\left(\sqrt{3}\cos{\varphi}\sin{\alpha}-3\cos{\alpha}\sin{\varphi}\right)}}$. Here, $g$ is a geometric factor that relates the sample geometry with the signal amplitude in the NV site, $k_B$ is the Boltzmann constant, and $T$ is the temperature. See \cite{Glenn18, Munuera23} for further details on the signal amplitude expression. It can be checked that $\varphi_{\rm opt}$ does indeed maximize $D_\varphi$. The total accumulated phase of the three drivings $A$, $\bar A$, $B$ is simply $3\Phi_k$, provided that $\nu_\varphi = \mu-\beta_\varphi$ is the same in the three stages, which in our case we choose to add up to $0$.

Finally, to consider all effective chemical shifts $\delta^*_i$ it suffices to sum all the contributions. Assuming a small angle $\Phi_k$, the final formula for the expected value of $\sigma_z$ is
\begin{equation}
    \langle\sigma_z\rangle_k \approx 3 D_\varphi\sum_i{\left[\rho_i\cos{\left(\frac{4\delta_i^* k}{\bar{\Omega}}+\nu_\varphi\right)}\right]},
\end{equation}
which gives us the desired spectrum upon Fourier transform.


\begin{thebibliography}{99} 
\bibitem{Dowling03} J. P. Dowling, and G. J. Milburn, { Quantum technology: the second quantum revolution}, \href{https://doi.org/10.1098/rsta.2003.1227}{Phil. Trans. R. Soc. A. {\bf361}, 1655 (2003).}

\bibitem{Degen17} C. L. Degen, F. Reinhard, and P. Cappellaro, { Quantum sensing}, \href{https://journals.aps.org/rmp/abstract/10.1103/RevModPhys.89.035002}{Rev. Mod. Phys. {\bf 89}, 035002 (2017).}

\bibitem{Levitt08} M. H. Levitt, { Spin Dynamics: Basics of Nuclear Magnetic Resonance}, 2nd ed. (Wiley, West Sussex, 2008).

\bibitem{Doherty13} M. W. Doherty, N. B. Manson, P. Delaney, F. Jelezko, J. Wrachtrup, and L. C. L. Hollenberg, { The nitrogen-vacancy colour centre in diamond}, \href{https://doi.org/10.1016/j.physrep.2013.02.001}{Phys. Rep. {\bf 528}, 1 (2013).}

\bibitem{Boss17} J. M. Boss, K. S. Cujia, J. Zopes, and C. L. Degen, { Quantum sensing with arbitrary frequency resolution}, \href{https://www.science.org/doi/10.1126/science.aam7009} {Science {\bf 356}, 837 (2017)}.

\bibitem{Schmitt17} S. Schmitt, T. Gefen, F. M. St\"urner, T. Unden, G. Wolff, C. M\"uller, J. Scheuer, B. Naydenov, M. Markham, S. Pezzagna, J. Meijer, I. Schwarz, M. B. Plenio, A. Retzker, L. P. McGuinness, and F. Jelezko, { Submillihertz magnetic spectroscopy performed with a nanoscale quantum sensor}, \href{https://www.science.org/doi/10.1126/science.aam5532} {Science {\bf 356}, 832 (2017)}.

\bibitem{Glenn18} D. R. Glenn, D. B. Bucher, J. Lee, M. D. Lukin, H. Park, and R. L. Walsworth, { High-resolution magnetic resonance spectroscopy using a solid-state spin sensor}, \href{https://doi.org/10.1038/nature25781}{Nature {\bf555}, 351 (2018)}.

\bibitem{Arunkumar21}N. Arunkumar, D. B. Bucher, M. J. Turner, P. TomHon, D. Glenn, S. Lehmkuhl, M. D. Lukin, H. Park, M. S. Rosen, T. Theis, and R. L. Walsworth, { Micron-Scale NV-NMR Spectroscopy with Signal Amplification by Reversible Exchange}, \href{https://journals.aps.org/prxquantum/abstract/10.1103/PRXQuantum.2.010305}{PRX Quantum {\bf2}, 010305 (2021)}.

\bibitem{Bucher20} D. B. Bucher, D. R. Glenn, H. Park, M. D. Lukin, and R. L. Walsworth, { Hyperpolarization-Enhanced NMR Spectroscopy with Femtomole Sensitivity Using Quantum Defects in Diamond}, \href{https://journals.aps.org/prx/abstract/10.1103/PhysRevX.10.021053}{Phys. Rev. X {\bf10}, 021053 (2020)}.

\bibitem{Ugurbil03}K. U\v gurbil, G. Adriany, P. Andersen, W. Chen, M. Garwood, R. Gruetter, P. G. Henry, S. G. Kim, H. Lieu, I. Tkac, T. Vaughan, P. F. Van De Moortele, E. Yacoub, and X. H. Zhu, { Ultrahigh field magnetic resonance imaging and spectroscopy}, \href{https://www.sciencedirect.com/science/article/pii/S0730725X03003382?via%3Dihub}{Magn. Reson. Imaging {\bf21}, 1263 (2003).}

\bibitem{Alsina23} P. Alsina-Bol\'ivar, A. Biteri-Uribarren, C. Munuera-Javaloy, and J. Casanova, { J-coupling NMR Spectroscopy with Nitrogen Vacancy Centers at High Fields}, \href{https://doi.org/10.48550/arXiv.2311.11880}{arXiv preprint:2311.11880 (2023).} 

\bibitem{Munuera23} C. Munuera-Javaloy, A. Tobalina, and J. Casanova, { High-Resolution NMR Spectroscopy at Large Fields with Nitrogen Vacancy Centers}, \href{https://doi.org/10.1103/PhysRevLett.130.133603}{Phys. Rev. Lett. {\bf130}, 133603 (2023).}

\bibitem{Meinel23} J. Meinel, M. Kwon, R. Maier, D. Dasari, H. Sumiya, S. Onoda, J. Isoya, V. Vorobyov, and J. Wrachtrup, { High-resolution nanoscale NMR for arbitrary magnetic fields}, \href{https://doi.org/10.1038/s42005-023-01419-2}{Commun. Phys. {\bf 6}, 302 (2023)}.

\bibitem{Daly23} D. Daly, S.J. DeVience, E. Huckestein, J.W. Blanchard, J. Cremer and R.L. Walsworth, { Nutation-Based Longitudinal Sensing Protocols for High-Field NMR With Nitrogen-Vacancy Centers in Diamond}, \href{https://arxiv.org/abs/2310.08499}{ arXiv preprint:2310.08499 (2023).}

\bibitem{Madhu16} K. R. Mote, V. Agarwal, and P. K. Madhu, { Five decades of homonuclear dipolar decoupling in solid-state NMR: Status and outlook,} \href{https://doi.org/10.1016/j.pnmrs.2016.08.001}{Prog. Nucl. Magn. Reson. Spectrosc. {\bf 97}, 1 (2016).}

\bibitem{Pisklak14} D. M. Pisklak, M. Zieli\'nska-Pisklak, L. Szeleszczuk, I. Wawer, { $^{13}$C cross-polarization magic-angle spinning nuclear magnetic resonance analysis of the solid drug forms with low concentration of an active ingredient-propranolol case}, \href{https://doi.org/10.1016/j.jpba.2013.06.031}{J. Pharm. Biomed. Anal. {\bf93}, 68 (2014).}

\bibitem{Pisklak16} D. M. Pisklak, M. A. Zieli\'nska- Pisklak, L. Szeleszczuk, I. Wawer, { $^{13}$C solid-state NMR analysis of the most common pharmaceutical excipients used in solid drug formulations, Part I: Chemical shifts assignment}, \href{https://doi.org/10.1016/j.jpba.2016.01.032}{J. Pharm. Biomed. Anal. {\bf122}, 81 (2016).}

\bibitem{Nie16} H. Nie, Y. Su, M. Zhang, Y. Song, A. Leone, L. S. Taylor, P. J. Marsac, T. Li, and S. R. Byrn, { Solid-State Spectroscopic Investigation of Molecular Interactions between Clofazimine and Hypromellose Phthalate in Amorphous Solid Dispersions}, \href{https://pubs.acs.org/doi/10.1021/acs.molpharmaceut.6b00740}{Mol. Pharmaceutics, {\bf13} 3964-3975 (2016).}

\bibitem{Marchetti21} A. Marchetti, J. Yin, Y. Su, X. Kong, { Solid-state NMR in the field of drug delivery: State of the art and new perspectives}, \href{https://doi.org/10.1016/j.mrl.2021.100003}{MRL {\bf 1}, 28-70 (2021).}

\bibitem{Tycko11} R. Tycko, { Solid State NMR Studies of Amyloid Fibril Structure}, \href{https://doi.org/10.1146/annurev-physchem-032210-103539}{Annu. Rev. Phys. Chem. {\bf 62}, 279-299 (2011).}

\bibitem{Pecher17} O. Pecher, J. Carretero-Gonz\'alez, K. J. Griffith, and C. P. Grey, { Materials' Methods: NMR in Battery Research}, \href{https://pubs.acs.org/doi/10.1021/acs.chemmater.6b03183}{Chem. Mater. {\bf 29}, 213-242 (2017).}

\bibitem{Allert22} R. D. Allert, K. D. Briegel, and D. B. Bucher, { Advances in nano- and microscale NMR spectroscopy using diamond quantum sensors}, \href{https://doi.org/10.1039/D2CC01546C}{Chem. Commun. {\bf 58}, 8165 (2022)}

\bibitem{Rizzato23} R. Rizzato, N. R. von Grafenstein, D. B. Bucher, { Quantum sensors in diamonds for magnetic resonance spectroscopy: Current applications and future prospects}, \href{https://doi.org/10.1063/5.0169027}{Appl. Phys. Lett. {\bf 123}, 260502 (2023)}.

\bibitem{Herb20} K. Herb, J. Zopes, K. S. Cujia, and C. L. Degen, { Broadband radio-frequency transmitter for fast nuclear spin control}, \href{https://doi.org/10.1063/5.0013776}{ Rev. Sci. Instrum. {\bf 91}, 113106 (2020)}.

\bibitem{Yudilevich23} D. Yudilevich, A. Salhov, I. Schaefer, K. Herb, A. Retzker, and A. Finkler, { Coherent Manipulation of nuclear spins in the strong driving regime}, \href{https://iopscience.iop.org/article/10.1088/1367-2630/ad0c0b}{ New. J. Phys. {\bf 25}, 113042 (2023}.

\bibitem{Lee65} M. Lee, and W. I. Goldburg, { Nuclear-Magnetic-Resonance Line Narrowing by a Rotating rf Field}, \href{https://journals.aps.org/pr/abstract/10.1103/PhysRev.140.A1261}{Phys. Rev. {\bf 140}, 1261 (1965)}.

\bibitem{Supp} See Supplemental material


\bibitem{Waugh72} M. Mehring, and J. S. Waught, { Magic-Angle NMR Experiments in Solids}, \href{https://doi.org/10.1103/PhysRevB.5.3459}{Phys. Rev. B {\bf 5}, 3459 (1972)}.


\bibitem{Vinogradov99} E. Vinogradov, P.K. Madhu, S. Vega, { High-resolution proton solid-state NMR spectroscopy by phase-modulated Lee-Goldburg experiment}, \href{https://doi.org/10.1016/S0009-2614(99)01174-4}{Chem. Phys. Lett. {\bf 314} 443 (1999)}. 


\bibitem{Halse13} M. E. Halse, and L. Emsley, { Improved Phase-Modulated Homonuclear Dipolar Decoupling for
Solid-State NMR Spectroscopy from Symmetry Considerations}, \href{https://pubs.acs.org/doi/10.1021/jp4038733} {J. Phys. Chem. A {\bf 117}, 5280 (2013)}.

\bibitem{supp1} See file single\_block.mp4 at (link).

\bibitem{supp2} See file global\_precession.mp4 at (link).




\bibitem{Purcell46} H. Y. Carr, and E. M. Purcell, { Effects of Diffusion on Free Precession in Nuclear Magnetic Resonance Experiments}, \href{https://doi.org/10.1103/PhysRev.94.630}{Phys. Rev. {\bf 94}, 630 (1954)}.

\bibitem{Meiboom58} S. Meiboom, and D. Gill, { Modified Spin-Echo Method for Measuring Nuclear Relaxation Times}, \href{https://doi.org/10.1063/1.1716296}{Rev. Sci. Instrum. {\bf 29}, 688-691 (1958)}.































\end{thebibliography}

\begin{thebibliography}{99} 

\bibitem{Glenn18} D. R. Glenn, D. B. Bucher, J. Lee, M. D. Lukin, H. Park, and R. L. Walsworth, { High-resolution magnetic resonance spectroscopy using a solid-state spin sensor}, \href{https://doi.org/10.1038/nature25781}{Nature {\bf555}, 351 (2018)}.

\bibitem{Munuera23} C. Munuera-Javaloy, A. Tobalina, and J. Casanova, { High-Resolution NMR Spectroscopy at Large Fields with Nitrogen Vacancy Centers}, \href{https://doi.org/10.1103/PhysRevLett.130.133603}{Phys. Rev. Lett. {\bf130}, 133603 (2023).}


\end{thebibliography}
\end{document}